**Magnetic and electrical transport signatures of uncompensated moments in epitaxial thin films of the non-collinear antiferromagnet Mn$_3$Ir**

James M. Taylor * [1], Edouard Lesne [1], Anastasios Markou [2], Fasil Kidane Dejene [1], Pranava Keerthi Sivakumar [1], Simon Pöllath [3], Kumari Gaurav Rana [1], Neeraj Kumar [1], Chen Luo [4][5], Hanjo Ryll [4], Florin Radu [4], Florian Kronast [4], Peter Werner [1], Christian H. Back [3][5], Claudia Felser [2], Stuart S. P. Parkin † [1]

[1] Max Planck Institute of Microstructure Physics, Weinberg 2, 06120 Halle (Saale), Germany
[2] Max Planck Institute for Chemical Physics of Solids, Nöthnitzer Str. 40, 01187 Dresden, Germany
[3] Institute of Experimental Physics, University of Regensburg, 93040 Regensburg, Germany
[4] Helmholtz-Zentrum Berlin for Materials and Energy, Albert Einstein Str. 15, Berlin 12489, Germany
[5] Institute of Experimental Physics of Functional Spin Systems, Technical University Munich, James-Franck-Str. 1, 85748 Garching b. München, Germany

**Abstract**

Non-collinear antiferromagnets, with either an $L1_2$ cubic crystal lattice (e.g. Mn$_3$Ir and Mn$_3$Pt) or a $D0_{19}$ hexagonal structure (e.g. Mn$_3$Sn and Mn$_3$Ge), exhibit a number of novel phenomena of interest to topological spintronics. Amongst the cubic systems, for example, tetragonally distorted Mn$_3$Pt exhibits an intrinsic anomalous Hall effect (AHE). However, Mn$_3$Pt only enters a non-collinear magnetic phase close to the stoichiometric composition and at suitably large thicknesses. Therefore, we turn our attention to Mn$_3$Ir, the material of choice for use in exchange bias heterostructures. In this paper, we investigate the magnetic and electrical transport properties of epitaxially grown, face-centered-cubic γ-Mn$_3$Ir thin films with (111) crystal orientation. Relaxed films of 10 nm thickness exhibit an ordinary Hall effect, with a hole-type carrier concentration of $(2.24 \pm 0.08) \times 10^{23}$ cm$^{-3}$. On the other hand, TEM characterization demonstrates that ultrathin 3 nm films grow with significant in-plane tensile strain. This may explain a small remanent moment, observed at low temperatures, shown by XMCD spectroscopy to arise from uncompensated Mn spins. Of the order 0.02 μ$_B$ / atom, this dominates electrical transport behavior, leading to a small AHE and negative magnetoresistance. These results are discussed in terms of crystal microstructure and chiral domain behavior, with spatially resolved XML(C)D-PEEM supporting the conclusion that small antiferromagnetic domains, < 20 nm in size, of differing chirality account for the absence of observed Berry curvature driven magnetotransport effects.

---

* james.taylor@mpi-halle.mpg.de
† stuart.parkin@mpi-halle.mpg.de

Antiferromagnetic (AF) spintronics is a growing research field [1], motivated by a number of potential advantages for applications including ultrafast magnetization dynamics [2] and improved stability against external perturbations at reduced dimensions. For example, synthetic antiferromagnetic structures (SAFs) [3] are already employed to eliminate magnetostatic fields in spin-valve sensors [4] and magnetic tunnel junction memory bits [5]. Furthermore, an efficient current-driven domain-wall motion has been demonstrated in such SAFs [6] by utilizing the chirality of the magnetic structure [7].

Indeed, chiral spin textures play a key role in the emerging field of topological AF spintronics [8]. Of particular interest are the non-collinear AFs $Mn_3X$ ($X$ = Ir, Pt, Sn, Ge), which can stabilize with either a face-centered-cubic (fcc) or a hexagonal crystal structure. The symmetry breaking non-collinear spin texture of these materials, combined with spin-orbit coupling, gives rise to a Berry curvature driven effective field, that is predicted to generate an intrinsic anomalous Hall effect (AHE) [9,10]. In the case of cubic $Mn_3Ir$, a facet-dependent spin Hall effect (SHE) emerging from the same origin has been discovered [11]. Whilst this intrinsic SHE is even with respect to the handedness of topological AF order, contributions to the AHE will cancel out over domains with opposite chirality of spin texture [12].

Experimental measurements of the intrinsic AHE have been realized in bulk single crystals of hexagonal non-collinear AFs, namely $Mn_3Sn$ [13] and $Mn_3Ge$ [14]. This has been enabled by the small in-plane magnetic moment exhibited by these materials, also demonstrated in epitaxial thin films of $Mn_3Sn$ [15], which arises from geometric frustration of the inverse triangular spin texture. Alignment of this weak magnetization via an external magnetic field in turn coherently orientates the AF order throughout the material, driving it into a dominant chiral domain state [16].

Such control of AF domains is challenging in the fcc non-collinear systems, which are normally fully compensated with strong internal anisotropy fields [17]. However, Liu *et al.* [18] have recently discovered a large AHE in $Mn_3Pt$ films epitaxially grown with in-plane tensile strain on $BaTiO_3$ substrates. $Mn_3Pt$ undergoes a first-order magnetic phase transition from the non-collinear to a collinear AF state above ≈ 360 K. AHE is only observed below this transition temperature, where a small uncompensated magnetization is also measured. The size of this net moment correlates with the magnitude of Berry curvature driven AHE and shows a dependence on the degree of tetragonal distortion of the films, although the exact relationship between them (including the role of chiral domain manipulation) remains open to further study.

Whilst the magnetic phase transition of Mn$_3$Pt has allowed the important demonstration of electric field control of this topological AHE, by applying additional piezoelectric strain to move between the collinear and non-collinear states [18], it also presents an upper limit to the operating temperature in resulting spintronic applications. On the other hand, the closely related cubic non-collinear AF Mn$_3$Ir can be stabilized with a triangular spin texture below its high Néel temperature, $T_N$ = 700 K [19], over a broad composition range in the phase diagram [20].

In this letter, the role of Mn$_3$Ir evolves to become the active element of potential future chiral spintronic devices, as we explore further the subtle interplay between crystal microstructure, uncompensated moments and electrical transport properties in this non-collinear AF.

In order to elucidate the behavior of topologically driven phenomena in these materials, thin film samples with high-quality crystal structure are required [21]. The epitaxial films utilized here were grown by magnetron sputtering according to our recipe published in Ref. [22]. Mn$_3$Ir films with (111) planes parallel to the substrate surface were selected for further study, grown with sample structure: Al$_2$O$_3$ (0001) [Substrate] / TaN (111) [5 nm] / Mn$_{(0.72 \pm 0.03)}$Ir$_{(0.28 \pm 0.03)}$ (111) [3 or 10 nm] / TaN [2.5 nm]. The two different sample thicknesses were chosen to display different structural properties, whilst both having a $T_N$ above room temperature (RT) [23]. Comprehensive characterization is detailed in Ref. [22], demonstrating that these films grow in an fcc γ-Mn$_3$Ir phase and suggesting a non-collinear magnetic structure [19].

As detailed in the Supplementary Material, 10 nm Mn$_3$Ir (111) films grow fully relaxed, with a large laterally-oriented grain size ≥ 20 nm. In the case of 3 nm ultrathin films, high-resolution TEM operated at 300 kV (FEI Titan 80-300) was used to further analyze crystal structure. Fig. 1(a) shows a typical micrograph of the epitaxial growth of <111> oriented Mn$_3$Ir, with sharp interfaces and uniform thickness. Examples of grain boundary defects are highlighted, indicating a grain size in the lateral direction of 15 to 20 nm. The inset of Fig. 1(a) displays a fast Fourier transform diffractogram, taken from the marked area. Indexing of Mn$_3$Ir (111) and (002) diffraction peaks allows the calculation of out-of-plane, OP ($d_{111}$), and in-plane, IP ($d_{\bar{1}\bar{1}2}=\sqrt{2/3}\,d_{002}$), lattice spacing respectively. In the OP direction, the lattice plane separation, $d_{111}$ = (2.19 ± 0.09) Å, agrees within uncertainty with the bulk value (2.182 Å). Meanwhile, the estimated IP crystal lattice spacing, $d_{\bar{1}\bar{1}2}$ = (1.9 ± 0.1) Å, is significantly larger than the bulk (1.543 Å), indicating the ultrathin Mn$_3$Ir (111) films grow with appreciable IP strain.

Characterization of these samples' magnetic properties using SQUID vibrating sample magnetometry proved challenging, as explained in the Supplementary Material. Instead, direct measurements of the films' magnetic moment ($m$) were performed using X-ray magnetic circular dichroism (XMCD) spectroscopy at the VEKMAG endstation of the PM2 beamline at BESSY [24]. X-ray absorption spectra (XAS) were recorded around the Mn-$L_3$ edge in total electron yield mode (shown in the inset of Fig. 1(b)), using alternating right- and left-circularly polarized X-rays ($\sigma^+$ and $\sigma^-$ respectively), at a temperature $T$ = 10 K. Fig. 1(b) displays the resulting XMCD signal, ($\sigma^+$-$\sigma^-$)/($\sigma^+$+$\sigma^-$), as a function of OP applied magnetic field, $\mu_0 H \parallel$ [111]. In the case of a 10 nm $Mn_3Ir$ (111) film, a small linear response of Mn magnetic moment (calculated using the XMCD sum rules [25]) is observed, explained by their slight canting out of the (111) plane under the influence of an external magnetic field. Due to the high magnetic anisotropy of $Mn_3Ir$ [17], we estimate that the fields used during these experiments remain well below any spin-flip transition.

For a 3 nm $Mn_3Ir$ (111) film, an XMCD signal which is hysteretic for $\mu_0 H$ < 2 T is measured, with a remanent signal of around 0.5% (corresponding to a Mn moment of 0.02 $\mu_B$ / atom) and a coercivity of approximately 0.3 T. This demonstrates a small net magnetization in ultrathin $Mn_3Ir$ (111) films, which can be manipulated by an external magnetic field and arises from uncompensated Mn spins. Since both chemical composition and defect density are similar in 10 nm and 3 nm thick samples, this effect could be interface driven, for example, by a re-orientation of non-collinear AF structure [26]. However, no similar effect has been observed in ultrathin polycrystalline $Mn_{0.8}Ir_{0.2}$ films [27]. Therefore, the origin of this uncompensated moment may instead be strain in our epitaxial 3 nm $Mn_3Ir$ (111) films, consistent with the results of Liu *et al.* [18].

The thin films were then patterned into Hall bars, using electron beam lithography and Ar ion etching, with dimensions ranging from 150 × 50 µm$^2$ down to 3 × 1 µm$^2$. Current ($I_C$ = 200 µA) flow was directed along Hall bars fabricated in different IP crystalline directions. The inset of Fig. 2(b) displays the electrical measurement geometry. Previous magnetotransport measurements in $Mn_{0.8}Ir_{0.2}$ utilized anisotropic magnetoresistance (AMR) [28] or tunneling-AMR [29] to detect AF order. In our case, longitudinal ($\rho_{xx}$) and transverse ($\rho_{xy}$) resistivity were measured at different $T$, as a function of external magnetic field applied OP.

Fig. 2(a) records measurements of transverse resistivity in a 75 × 25 µm$^2$ Hall bar of a 3 nm $Mn_3Ir$ (111) film. We observe an anomalous-type behavior of $\rho_{xy}$ at 2 K, saturating at comparable fields to the hysteretic part of the XMCD signal. The inset of Fig. 2(a) shows this is accompanied by a negative longitudinal MR, ($[\rho_{xx}(\mu_0H)-\rho_{xx}(0)]/\rho_{xx}(0)$)×100%, which points to a magnetic origin, namely the presence of uncompensated Mn moments, for the exhibited AHE in strained ultrathin $Mn_3Ir$ (111) films. Similar electrical measurements of uncompensated Mn spins have been made at RT by Kosub *et al.* [30]. In our case, measurements in Fig. 2(a) at 50 K show both AHE and negative MR disappearing, indicating the strain-induced net Mn magnetization is sensitive to thermal fluctuations. In spite of the vanishing of remanent Mn moment at higher temperatures, a lack of change to transport properties seen in Fig. 2(a) after cooling from 400 K in a 9 T magnetic field ‖ [111] implies that samples nevertheless remain antiferromagnetic up to and above RT (i.e. have respective $T_N$ > 400 K), because it is known that field cooling analogous $Mn_{0.8}Ir_{0.2}$ films through $T_N$ can modify their electrically-detected AF order [31].

Fig. 2(b) shows transverse resistivity in 75 × 25 µm$^2$ Hall bars of a 10 nm $Mn_3Ir$ (111) film at 2 K. A positive linear response of $\rho_{xy}$ is measured, indicating dominant hole-type charge carriers. Fitting the gradient ($\rho_{xy}/\mu_0H$) of this ordinary Hall effect allows determination of the charge carrier concentration, $h = (2.24 \pm 0.08) \times 10^{23}$ cm$^{-3}$. In an attempt to clarify the absence of Berry curvature driven AHE, $\rho_{xy}$ was measured in Hall bar devices fabricated along different crystalline directions, because the intrinsic AHE is predicted to by highly anisotropic [13]. However, a linear Hall effect is observed along all crystallographic axes, displayed in Fig. 2(b) for the examples of the [$\bar{1}\bar{1}2$] and [$1\bar{2}1$] directions. This isotropic behavior suggests any intrinsic AHE may cancel over multiple degenerate AF domains (with opposing orientation of triangular spin texture) which follow the six-fold symmetry of the epitaxial crystal structure. To isolate individual chiral domains, we performed measurements in narrow channel Hall bars down to 1 µm in width, shown in Fig. 2(b) for an exemplar 15 x 5 µm$^2$ device. Again, the presence of a dominating ordinary Hall effect observed in all devices points to AF domain size in our epitaxial $Mn_3Ir$ (111) thin films being significantly smaller than the lowest device dimension tested (1 µm).

Therefore, in an attempt to elucidate the chiral domain structure of 10 nm $Mn_3Ir$ (111) thin films, we performed X-ray magnetic linear (circular) dichroism photo-emission electron microscopy, XML(C)D-PEEM (or X-PEEM), at beamline UE49_PGM at BESSY. Experimental details are described in the Supplementary Material. Fig. 3(a) shows an XMCD-PEEM image, taken at the Mn-L$_3$ edge, at 45 K with no applied external magnetic field. For this film without remanent Mn moment, no XMCD contrast is observed above the sample surface topography background.

XMLD-PEEM imaging has been shown to exhibit contrast between domains with orthogonal Néel vector orientations in collinear AFs [32,33]. We postulate that, in the same way, differences in orientation between the linearly polarized X-rays and the Néel vector defining the chirality of the triangular spin texture would lead to a difference in absorption between opposite chirality AF domains. Fig. 3(b) shows such an XMLD-PEEM image measured at the Mn-$L_3$ edge for the same 10 nm $Mn_3Ir$ (111) sample; no XMLD contrast is discerned. Possible reasons are discussed in the Supplementary Material, one of which may be that chiral domains are smaller than the resolution limit of the PEEM microscope (≈ 20 nm), which will indeed be the case if they are correlated with the grain size in the film measured using TEM.

Finally, in an attempt to enlarge chiral domains in $Mn_3Ir$ to an observable size, exchange bias was utilized to introduce a preferential AF domain orientation through coupling to a ferromagnetic (FM) layer [34]. X-PEEM was therefore imaged at both the Ni-$L_3$ and Mn-$L_3$ edges in a 3nm $Mn_3Ir$ (111) / 5 nm $Ni_{80}Fe_{20}$ bilayer. Fig. 3(c) exhibits an approximately equal distribution of oppositely oriented FM domains in an XMCD-PEEM image recorded at the Ni-$L_3$ edge at RT. After cooling the bilayer to 70 K under a 20 mT IP magnetic field, Fig. 3(d) displays a repeat XMCD-PEEM image of the same area in which the FM domains have grown, but no preferential domain direction has been set. This is likely due to the bilayer not having passed through its blocking temperature of 40 K [22]. Finally, XMCD- and XMLD-PEEM images were recorded at the Mn-$L_3$ edge after this IP field cooling routine, shown in Figs. 3(e) and (f) respectively. No uncompensated Mn spins are observed at the interface, as expected if the temperature is not low enough to induce large exchange bias [35]. Finally, no AF domains of differing chirality are resolved, which may be due to the concomitant difficulty of observing the buried interface through a 5 nm $Ni_{80}Fe_{20}$ layer [36] combined with intrinsic spatial resolution limit of X-PEEM.

In conclusion, we studied the magnetic and electrical properties of fully relaxed 10 nm $Mn_3Ir$ (111) samples, and of ultrathin films exhibiting significant IP lattice distortion. This tensile strain may be the origin of an uncompensated Mn magnetic moment observed by XMCD spectroscopy. Because of this net Mn magnetization, the 3 nm $Mn_3Ir$ (111) films demonstrate a small negative MR and AHE at low temperature. On the other hand, 10 nm $Mn_3Ir$ (111) films exhibit ordinary Hall effect, which is isotropic with respect to IP crystallographic direction. A lack of Berry curvature driven electrical transport effects can be explained by the presence of multiple AF domains of differing triangular spin texture chirality, suggested by XML(C)D-PEEM imaging to be correlated with film grain size and thus < 20 nm. Our results illuminate the intimate connection between crystal structure, uncompensated spins and magnetotransport properties, therefore informing the further implementation of non-collinear Mn$_3$X thin films in *chiralitronic* devices.

**Supplementary Material**

See supplementary material for summarized characterization of 10 nm Mn$_3$Ir (111) thin films, and further details of magnetic measurements and X-PEEM experimental procedures.

**Acknowledgments**

We acknowledge Helmholtz-Zentrum Berlin (HZB) for synchrotron beamtime at UE49_PGM SPEEM (proposal 181-06569) and at PM2 VEKMAG (proposal 171-04784). Financial support for the VEKMAG project and for the PM2-VEKMAG beamline are provided by the German Federal Ministry for Education and Research (BMBF 05K10PC2, 05K10WR1, 05K10KE1) and by HZB, with Steffen Rudorff thanked for technical support. This work was partially funded by ASPIN (EU H2020 FET open grant 766566).

**FIG. 1.** (a) TEM image of a 3 nm $Mn_3Ir$ (111) film, viewed along the $[1\bar{1}0]$ zone axis, with in-plane (IP) crystallographic directions indicated and grain boundaries highlighted by dashed lines (inset shows diffractogram from region marked by green box). (b) XMCD measured at 10 K for $Mn_3Ir$ (111) films with different thicknesses (inset shows XAS spectra recorded at the Mn-$L_3$ edge using right- and left-circularly polarized X-rays after sweeping magnetic field to -8 T, and the resulting XMCD spectrum).

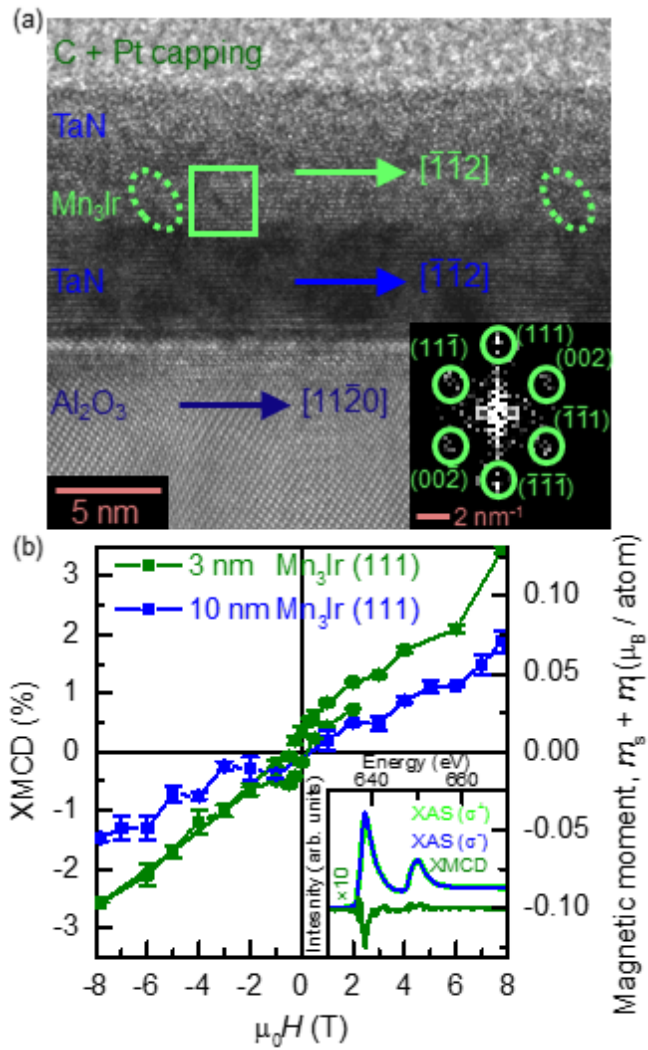

**FIG. 2.** (a) Hall effect for a 3 nm Mn$_3$Ir (111) film at 2 K, with and without 9 T OP field cooling, and at 50 K (inset shows variation in magnetoresistance at 2 K and 50 K). (b) Hall effect for a 10 nm Mn$_3$Ir (111) film at 2 K, in Hall bars fabricated along different crystallographic axes and of different dimensions (inset shows an optical image of example patterned device with measurement geometry indicated).

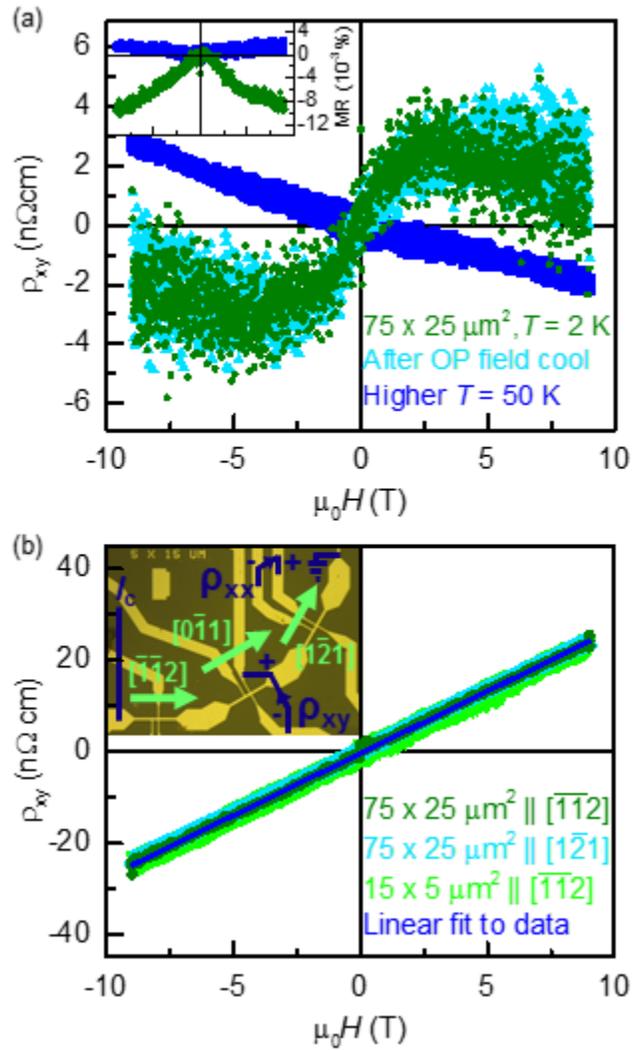

**FIG. 3.** XPEEM images (5 × 5 µm$^2$) measured in a 10 nm Mn$_3$Ir (111) film using (a) XMCD and (b) XMLD at the Mn-L$_3$ edge; and in a 3 nm Mn$_3$Ir (111) / 5 nm Ni$_{80}$Fe$_{20}$ bilayer using XMCD at the Ni-L$_3$ edge both at (c) 300 K and (d) 70 K after 20 mT IP field cooling, and using (e) XMCD and (f) XMLD at the Mn-L$_3$ edge after 20 mT IP field cooling.

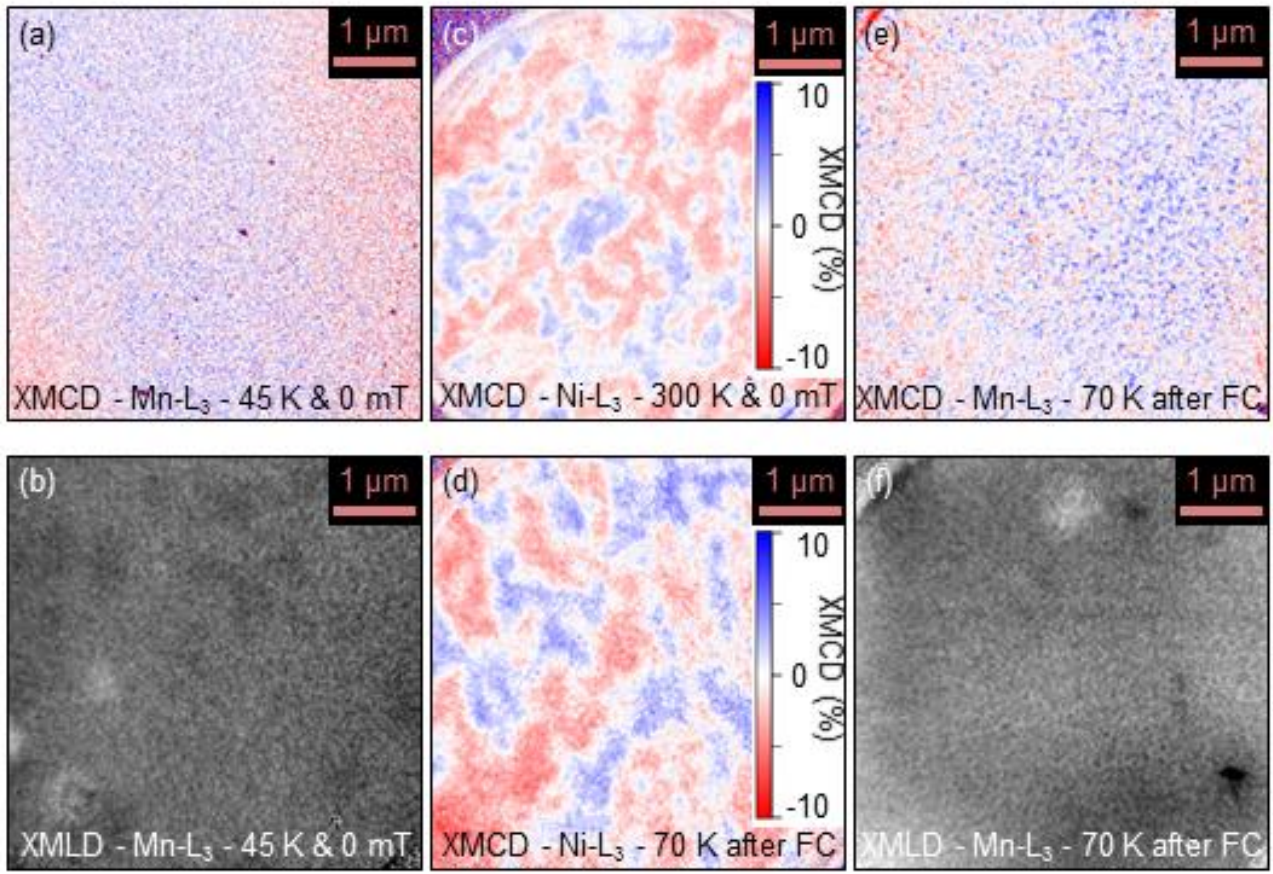

## Supplementary Material

### I – Growth and structural characterization

Comprehensive details of the thin film growth procedure and subsequent structural characterization are given in our previous paper [1]. These demonstrate that the <111> oriented films grow in an fcc γ-$Mn_3Ir$ phase ($Fm\bar{3}m$, space group = 225), shown by Kohn *et al.* to possess a non-collinear antiferromagnetic (AF) structure as detailed in Ref. [2]. Atomic force microscopy measurements indicate the films' smooth grown, with RMS roughness < 4 Å. Meanwhile X-ray diffraction (XRD) ϕ scan pole figures confirm the pseudo hexagon-on-hexagon growth of the TaN (111) buffer, and in turn the $Mn_3Ir$ (111) thin film, on the $Al_2O_3$ (0001) substrate with the epitaxial relationship: $Al_2O_3$ (0001) [11$\bar{2}$0] [$\bar{1}$100] ‖ TaN (111) [$\bar{1}\bar{1}$2] [1$\bar{1}$0] ‖ $Mn_3Ir$ (111) [$\bar{1}\bar{1}$2] [1$\bar{1}$0].

In the main text, we determine the crystal lattice spacing for an ultrathin $Mn_3Ir$ film, and relate the observed in-plane (IP) tensile strain to the resulting magnetic properties. For the case of a thicker 10 nm $Mn_3Ir$ (111) film, Supplementary Fig. S1(a) displays a specular 2Θ-ω XRD scan, from which an out-of-plane (OP) lattice parameter of $c$ = (3.797 ± 0.001) Å is measured. Combined with transmission electron microscopy (TEM) studies reported in Ref. [1], such 10 nm thick films were found to grow fully relaxed, with both IP and OP lattice parameters close to the bulk value ($a$ = 3.780 Å). IP grain size > 20 nm is observed, indicating the growth of large crystallites in the lateral direction (slightly larger than in the ultrathin films). Further rocking curve measurements, ω XRD scans, an example of which is inset in Supplementary Fig. S1(a), show the films' low mosaicity of 0.5°. The combination of low roughness and mosaicity, with large lateral grain size and sharp crystal texture, demonstrate the high-quality epitaxial growth of these thin films.

### II – Magnetic properties

Initially, the magnetic moments of the 10 nm and 3 nm $Mn_3Ir$ (111) thin films were measured as a function of external magnetic field applied OP along the [111] direction ($\mu_0 H$) using the vibrating sample mode of a SQUID magnetometer (Quantum Design MPMS3). Supplementary Fig. S1(b) records such magnetic hysteresis loops for both 10 nm and 3 nm $Mn_3Ir$ (111) samples at room temperature (RT) and 5 K. In addition, measurements of a reference $Al_2O_3$ substrate prepared under the same conditions, and of an ultrathin film after field cooling from 400 K in a 9 T magnetic field || [111], are shown. All samples exhibit a large paramagnetic signal, dominated by the substrate background, combined with a small non-linear low-field contribution (also present in the substrate control sample). These results demonstrate the futility of such measurements using conventional magnetometry applied to AF thin films, because of the challenge disentangling tiny film signals from magnetic contamination of the substrate during sample preparation and other background signals. Instead, element specific measurements of the magnetic moment arising from the metallic film alone were made using X-ray magnetic circular dichroism (XMCD) in total electron yield mode, as described in the main text.

## III – X-ray magnetic linear (circular) dichroism photo-emission electron microscopy, XML(C)D-PEEM (or X-PEEM)

In an attempt to elucidate AF domain structure, X-PEEM measurements were made in 10 nm $Mn_3Ir$ (111) thin films, and 3 nm $Mn_3Ir$ (111) / 5 nm $Ni_{80}Fe_{20}$ heterostructures. Further details of the bilayer samples are given in Ref. [1]. Experiments were performed at beamline UE49_PGM at BESSY, using a combination of linearly and circularly polarized X-ray radiation to excite photo-emitted electrons from the thin film surface, which are then spatially imaged using an electron microscope. For XMCD, X-PEEM images were formed by taking the difference of images recorded using left- and right-handed circularly polarized X-rays. Here the presence of ferromagnetic (FM) domains or uncompensated AF moments would be expected to produce non-zero XMCD signal, with spatial variation in orientation yielding contrast in the image. This is represented in the color scale of the XMCD-PEEM images in Fig. 3 of the main text, with blue (red) contrast showing areas of positive (negative) XMCD signal respectively. In images (c) and (d), XMCD at the Ni-$L_3$ edge is set on a scale of ± 10%, whilst in images (a) and (e) the scale bar is normalized to the maximum/minimum signal at the Mn-$L_3$ edge.

For XMLD measurements, linearly polarized X-rays were incident at an angle of 16° to the sample surface, with polarization lying in the film plane. Previous seminal studies have utilized the contrast produced by differing absorption between AF domains with Néel vectors aligned either parallel or perpendicular to the linear polarized X-rays, in order to image orthogonally oriented domains in collinear AFs [3,4]. Two images were taken with X-ray energies ($E$) tuned to both the absorption peak at the Mn-$L_3$ edge ($E_B$) and to a lower energy just below the rising edge of the peak ($E_A$), with the asymmetry between the images' signal intensities ($I$) giving XMLD contrast according to $[I_{Max}(E_B)-I_{Min}(E_A)]/[I_{Max}(E_B)+I_{Min}(E_A)]$ [5]. The greyscale in the XMLD-PEEM images in Fig. 3 of the main text corresponds to the magnitude of this asymmetry, with the maximum/minimum of the scale bar in images (b) and (f) set to give optimum contrast. We hypothesized that differences in orientation between the linearly polarized X-rays and the Néel vector defining the chirality of AF domains would lead to a difference in absorption across domains of opposite chirality, and hence a contrast in the photo-emitted electron image. However, as shown in Fig. 3(b) of the main text, no discernable XMLD contrast is observed for a 10 nm $Mn_3Ir$ (111) thin film, with no structure between topological domains of different chirality visible. We postulate this may be due to either the limited escape depth of photo-electrons from these films (that are capped with a 2.5 nm layer of TaN), AF domains in this high-anisotropy material being smaller than the resolution limit of the PEEM microscope (< 20 nm, which maybe be the case if domain size is correlated with grain size in the film), or the possible insensitivity of the triangular AF structure to in-plane polarized X-rays [6].

In an attempt to enlarge chiral domains in $Mn_3Ir$ to an observable size, exchange bias was utilized to introduce a preferential AF domain orientation through coupling to a FM layer. By cooling such AF / FM heterostructures in an external magnetic field, a unidirectional exchange anisotropy, and hence a dominant pinned AF domain state, can be set. Using a combination of XMLD- and XMCD-PEEM, a direct correlation between externally manipulatable FM domains and exchange coupled AF domains has previously been observed [6-8]. We therefore took X-PEEM images at both the Ni-$L_3$ and Mn-$L_3$ edges in a 3 nm $Mn_3Ir$ (111) / 5 nm $Ni_{80}Fe_{20}$ bilayer, as reported in the main text.

**Supplementary References**

**Supplementary FIG. S1.** (a) XRD 2Θ-ω pattern for a 10 nm $Mn_3Ir$ (111) thin film, with diffraction peaks indexed (inset shows XRD ω rocking curve about $Mn_3Ir$ (111) peak). (b) Magnetic moment measured at RT and 2 K for $Mn_3Ir$ (111) films with different thicknesses, for a 3 nm film after 7 T OP field cooling, and for a substrate reference sample.

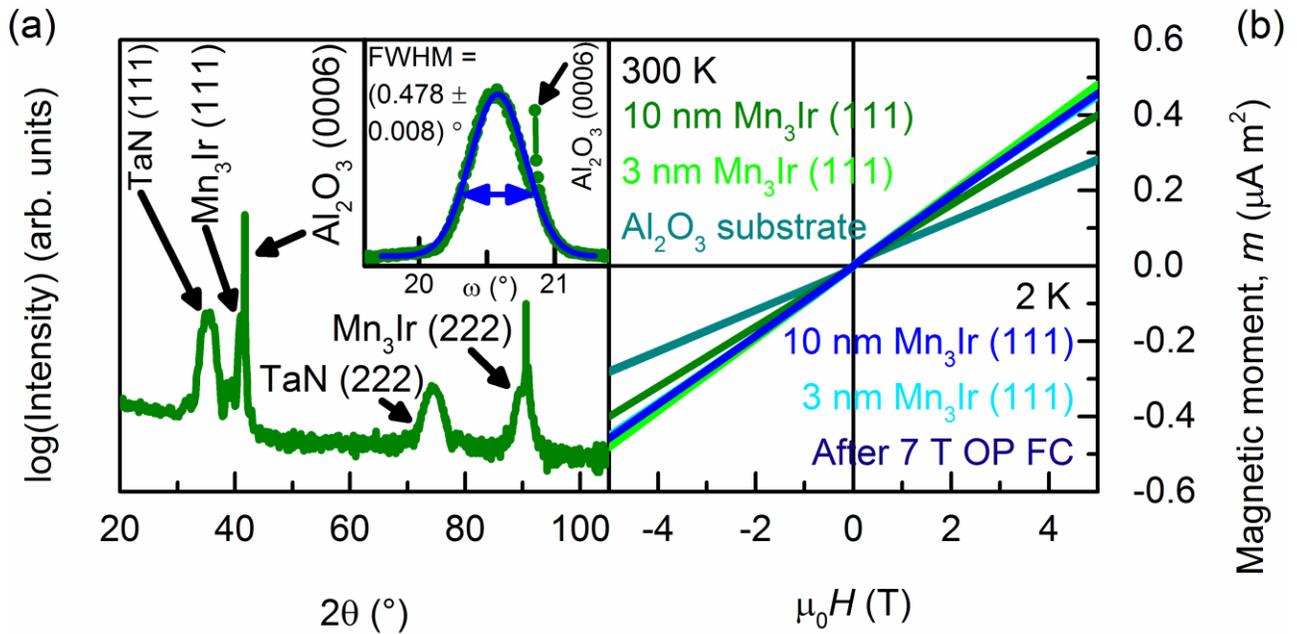